\documentclass[aps,prd,preprint,superscriptaddress]{revtex4}

\usepackage{graphicx}
\usepackage{amsmath}
\usepackage{bm}

\def\ee{\end{eqnarray}}

\def\=:{=\hspace{-.7em}\raisebox{1.1ex}{.}\hspace{.1em}\raisebox{-0.2ex}{.} }

\def\ee{\end{eqnarray}}

\def\=:{=\hspace{-.7em}\raisebox{1.1ex}{.}\hspace{.1em}\raisebox{-0.2ex}{.} }

\newcommand {\beq}{\begin{eqnarray}}
\newcommand {\eeq}{\end{eqnarray}}
\newcommand {\non}{\nonumber\\}

\newcommand {\1}[1]{\frac{1}{#1}}

\newcommand {\del}{\partial}

\begin{document}


\title{Matryoshka Skyrmions 
}


\author{Muneto Nitta}

\affiliation{
Department of Physics, and Research and Education Center for Natural 
Sciences, Keio University, Hiyoshi 4-1-1, Yokohama, Kanagawa 223-8521, Japan\\
}


\date{\today}
\begin{abstract}

We construct a stable Skyrmion in 3+1 dimensions 
as a sine-Gordon kink inside a domain wall within a domain wall 
in an $O(4)$ sigma model with hierarchical mass terms without the Skyrme term.
We also find that higher dimensional Skyrmions can stably exist with a help of non-Abelian domain walls in an $O(N)$ model with hierarchical mass terms without a Skyrme term, 
which leads to a matryoshka structure of Skyrmions.

\begin{center}
\includegraphics[width=0.4\linewidth,keepaspectratio]{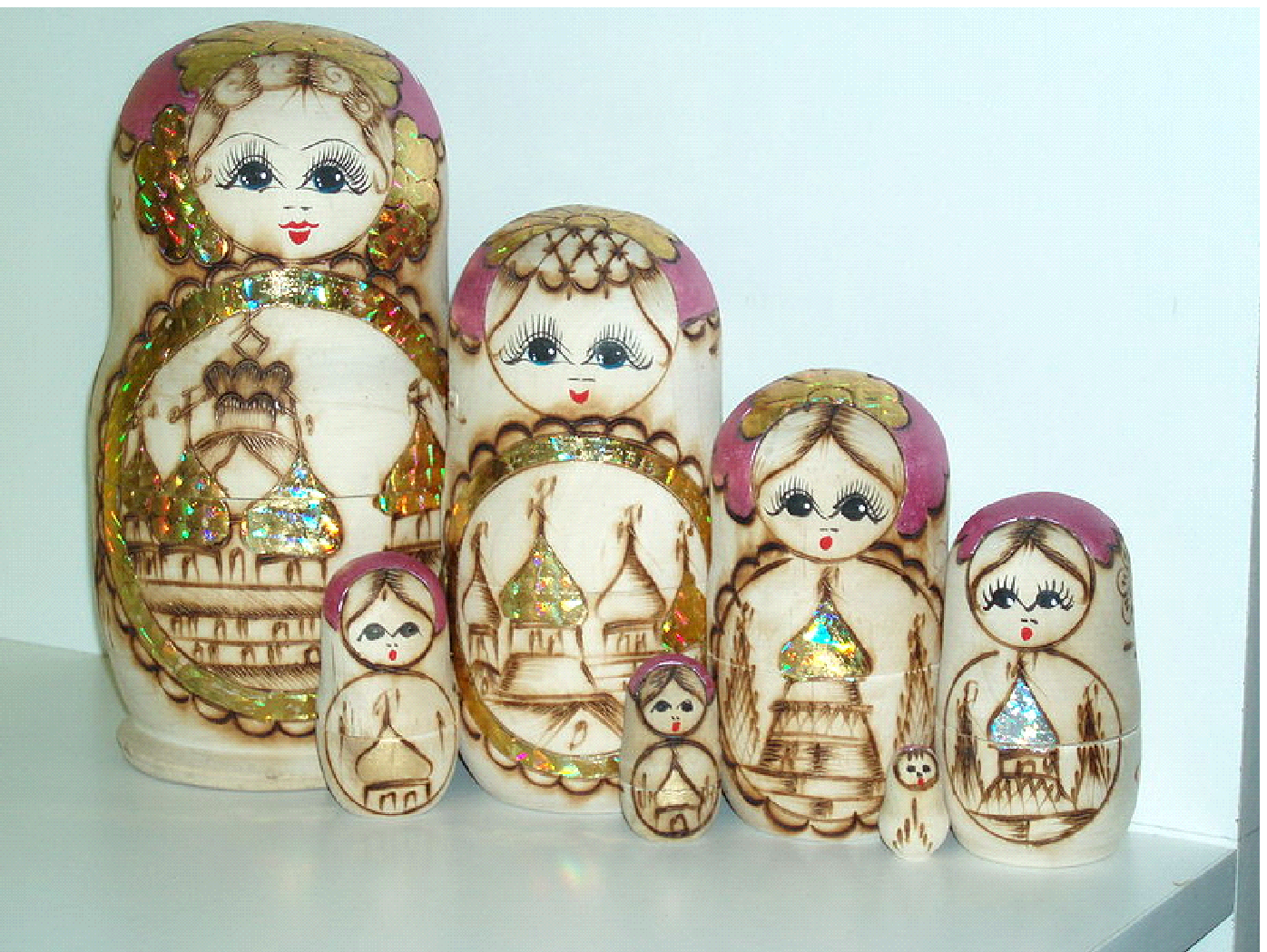}\\
a picture from Wikipedia
\end{center}

\end{abstract}

\maketitle

\section{Introduction}

A half century has been passing after
Skyrme proposed that 
Skyrmions characterized by the topological charge $\pi_3(S^3)\simeq{\bf Z}$ 
describe nucleons 
in the pion effective field theory or the chiral Lagrangian \cite{Skyrme:1962vh}, 
where the Skyrme term quartic with respect to derivatives is needed to 
stabilize Skyrmions against shrinkage. 
After Skyrmions were proposed, lower dimensional analogs of Skyrmions 
have also been studied  in both field theory and condensed matter physics.
In two spatial dimensions, 
a lump solution characterized by $\pi_2(S^2)\simeq{\bf Z}$ 
in an $O(3)$ sigma model \cite{Polyakov:1975yp} 
can be regarded as a two spatial dimensional (2D) Skyrmion, 
whose size is arbitrary.  
The size of lumps is fixed in the presence of 
a quartic derivative term analogous to the Skyrme term and a potential term. 
Such 2D Skyrmions were proposed as a toy model of Skyrmions \cite{Piette:1994ug} 
and are called baby Skyrmions.
A sine-Gordon kink characterized by $\pi_1(S^1)\simeq{\bf Z}$ 
is considered as the lowest dimensional Skyrmion,  
as was first studied by Skyrme \cite{Skyrme:1961vr}. 

In condensed matter physics, 
sine-Gordon kinks (1D Skyrmions) exist for instance in Josephson junctions of superconductors 
\cite{Ustinov:1998}.
2D Skyrmions appear in various condensed matter systems 
such as magnetism \cite{magnetism}, 
quatum Hall liquids \cite{QHE}, 
superfluid helium 3 \cite{Volovik:2003}, 
and spinor or multi-component Bose-Einstein condensates \cite{Leslie:2009}. 
More recently, there have been considerable efforts 
to realize 3D Skyrmions stably in 
two-component Bose-Einstein condensates \cite{3D-skyrmions}, 
which are elusive because of the lack of the Skyrme term.
Finally, it has been theoretically proposed that a stable 3D Skyrmion can exist 
as a ground state by introducing 
an ``artificial" non-Abelian gauge fields \cite{Kawakami:2012zw}.

As described above, 
Skyrmions exist in diverse dimensions \cite{Jackson:1988xk} 
in both field theory and condensed matter physics: 
an $O(N+1)$ nonlinear sigma model admits $N$ dimensional Skyrmions 
characterized by $\pi_{N}(S^{N})\simeq{\bf Z}$,
at least for $N=1,2,3$. 
However, it has been unclear thus far whether higher dimensional
Skyrmions with higher $N$ can exist stably.  
One of the purposes of this paper is to construct 
stable higher dimensional Skyrmions without a Skyrme term 
as composite states.

Relations between Skyrmions in diverse dimensions have been also studied.
A lump (2D Skyrmion) becomes a sine-Gordon kink (1D Skyrmion) 
 \cite{Kudryavtsev:1997nw,Auzzi:2006ju,Nitta:2012xq}, 
once it is absorbed 
into a ${\bf C}P^1$ domain wall \cite{Abraham:1992vb}.
This explains \cite{Nitta:2012xq}
a construction of a sine-Gordon kink from a holonomy of a ${\bf C}P^1$ lump \cite{Sutcliffe:1992ep}. 
A 3D Skyrmion becomes a 2D Skyrmion \cite{Kudryavtsev:1999zm,Nitta:2012wi}
once it is absorbed into a non-Abelian domain wall 
with $S^2$ moduli \cite{Losev:2000mm, Ritz:2004mp}.
The other purpose of this paper is to extend these relations 
between Skyrmions in diverse dimensions to arbitrary dimensions.

In this paper, we propose relations between Skyrmions in diverse dimensions through domain walls; 
when a $D$ dimensional Skyrmion is absorbed into a non-Abelian domain wall, 
it becomes a $D-1$ dimensional Skyrmion. 
By using this relation, we construct higher dimensional Skyrmions with a help of domain walls. 
These Skyrmions stably exist without the Skyrme term. 
More precisely, we consider an $O(N+1)$ nonlinear sigma model with the target space $S^{N}$ and 
a potential term with hierarchical masses.  
With the largest masses, there exists a non-Abelian domain wall 
 having normalizable moduli ${\bf R}\times S^{N-1}$, which 
connects two discrete vacua.
The domain wall effective action is thus an $O(N)$ model with the target space 
$S^{N-1}$ with a potential term with hierarchical masses. 
This effective Lagrangian admits again a non-Abelian domain wall 
with internal moduli ${\bf R}\times S^{N-2}$.
With repeating this procedure, we finally construct an $N$ dimensional Skyrmion 
as a sine-Gordon kink inside $N-1$ non-Abelian domain walls. 
In Sec.~\ref{sec:3dim}, we consider an $O(4)$ model in $d=3+1$ dimensions 
and construct a 3D Skyrmion as a sine-Gordon kink inside 
a domain wall within a domain wall. 
In Sec.~\ref{sec:Ndim}, we generalize this to higher dimensions 
and find a matryoshka structure of Skyrmions. 
Section \ref{sec:summary} is devoted to a brief summary and discussion. 

\section{O(4) model and 3D Skyrmions}\label{sec:3dim}

\subsection{A domain wall inside a domain wall}

\begin{figure}[ht]
\begin{center}
\begin{tabular}{cc}
\includegraphics[width=0.5\linewidth,keepaspectratio]{{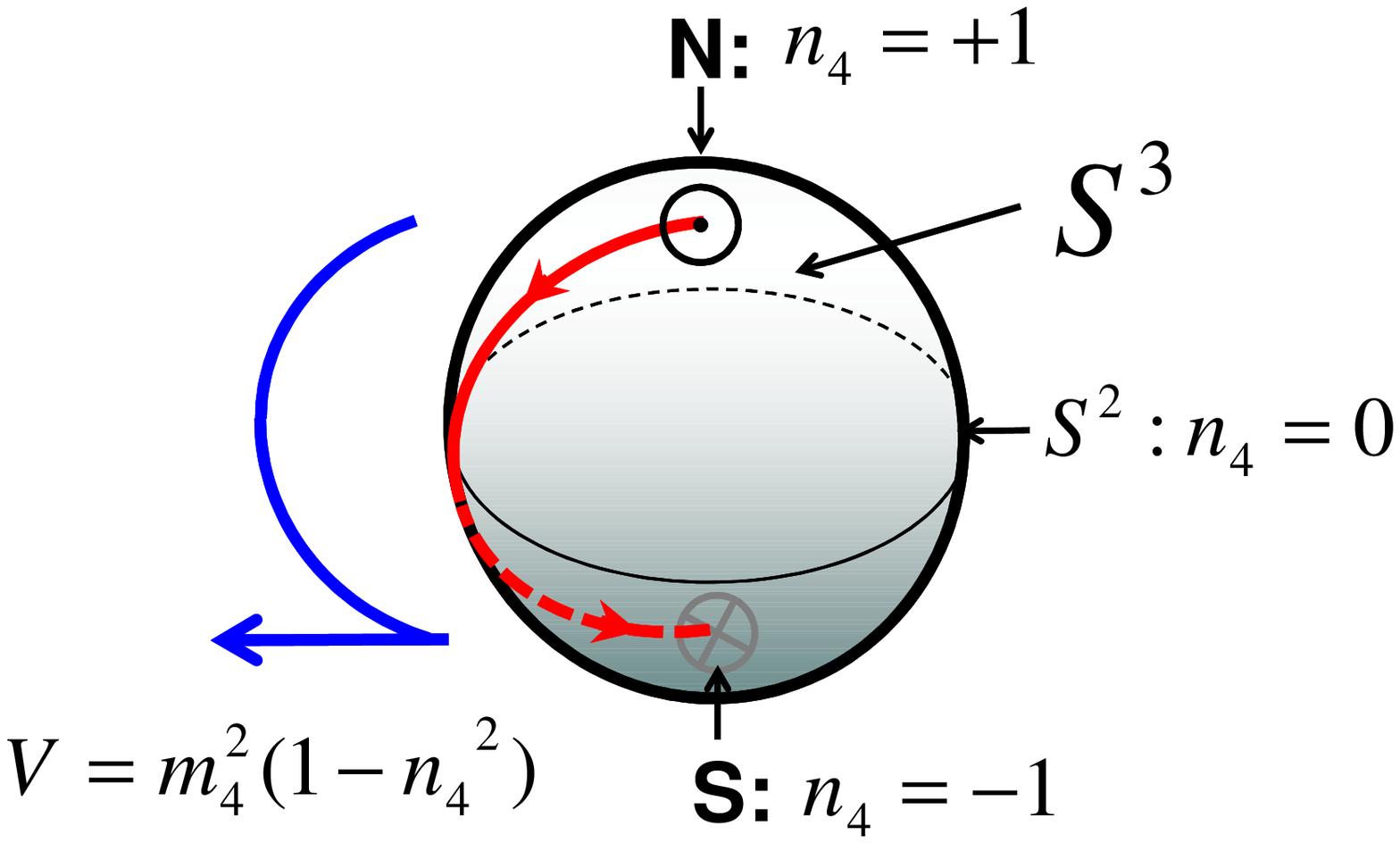}}
&
\includegraphics[width=0.5\linewidth,keepaspectratio]{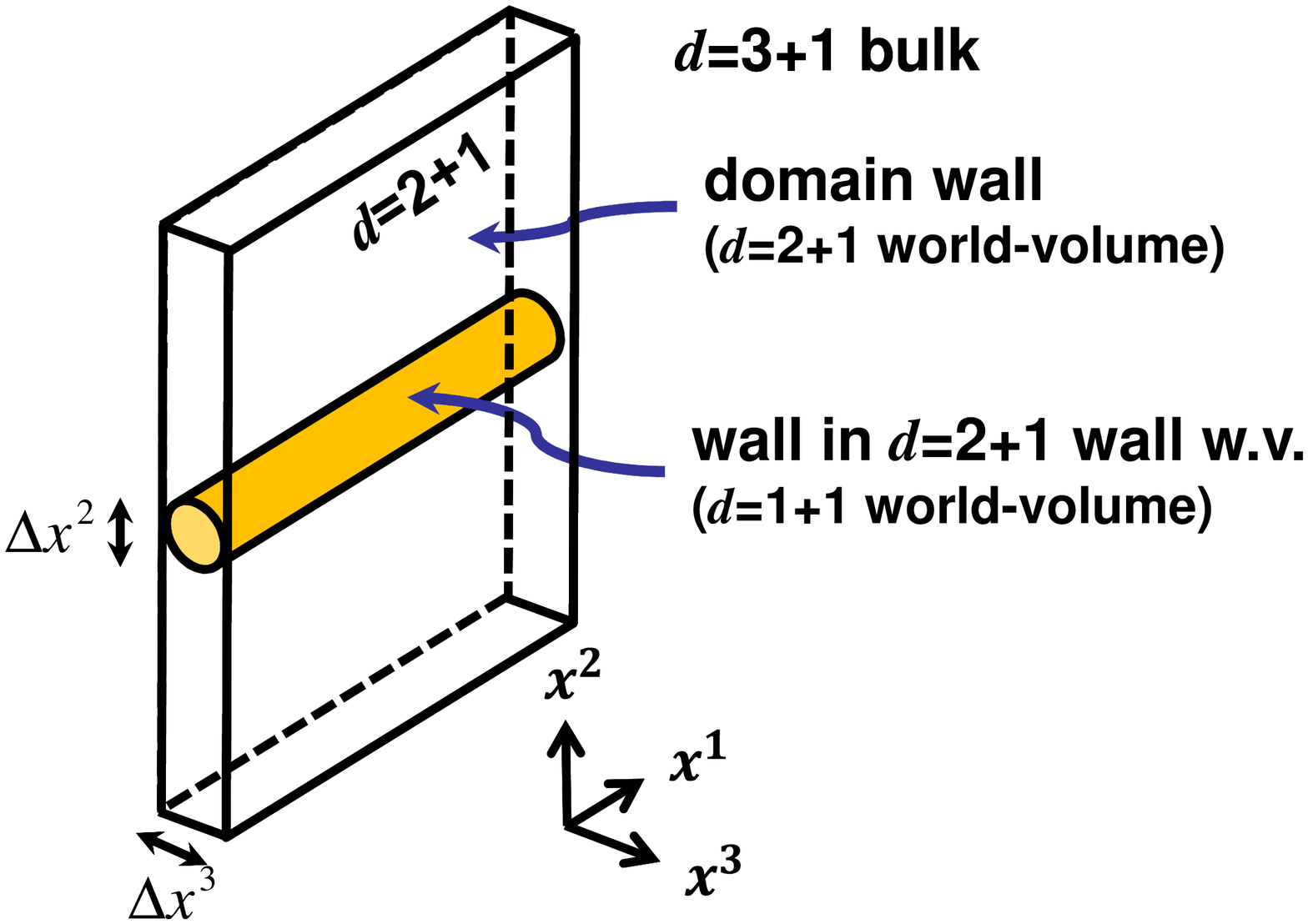}
\\
(a) & (b)
\end{tabular}
\caption{(a) The $S^3$ target space with a potential admitting two discrete vacua. 
(b) A wall within a wall. 
The widths of the outer and inner domain walls are $\Delta x^3 = m_4^{-1}$
and $\Delta x^2 = m_3^{-1}$, respectively. 
\label{fig:wall-wall}}
\end{center}
\end{figure}

Let ${\bf n}^{(0)}=(n^{(0)}_1(x),n^{(0)}_2(x),n^{(0)}_3(x),n^{(0)}_4(x))$ be an $O(4)$ vector of 
scalar fields satisfying the constraint $({\bf n}^{(0)})^2 =1$. 
By using these fields, 
the Lagrangian of an $O(4)$ nonlinear sigma model in $d=3+1$ dimensions can be given as 
($\mu=0,1,2,3$)
\beq
 {\cal L}^{(0)} = \1{2} (\partial_{\mu}{\bf n}^{(0)})^2 
  - V({\bf n}^{(0)}) .
\eeq 
Without the potential term, the Lagrangian is invariant under the $O(4)$ symmetry, 
which is spontaneously broken down to an $O(3)$ subgroup.  
Then, the target space is $O(4)/O(3) \simeq S^3 \simeq SU(2)$ [Fig.~\ref{fig:wall-wall}(a)], 
and this model is equivalent to the $SU(2)$ principle chiral model (or the chiral Lagrangian 
at the leading order). We do not consider the Skyrme term or other higher derivative terms.
Here, we consider the following potential term 
\beq 
V({\bf n}^{(0)}) = m_4^2\left(1-({n^{(0)}_4})^2\right)  + m_3^2 \left(1-({n^{(0)}_3})^2\right), \quad 
  m_3 \ll m_4, \label{eq:pot34}
\eeq
which hierarchically breaks the $O(4)$ symmetry explicitly to an $O(3)$ subgroup by $m_4$ 
and further to an $O(2)$ subgroup by $m_3$:
\beq
 O(4) \stackrel{m_4}{\to} O(3) \stackrel{m_3}{\to} O(2).
\eeq

When $m_3$ is negligible, the symmetry of vacua 
$n^{(0)}_4=\pm 1$ is $O(3)$, and
there exists a non-Abelian domain wall 
which interpolates these vacua by the field $n^{(0)}_4$ 
\cite{Losev:2000mm,Ritz:2004mp,Nitta:2012wi}.
We place the domain wall perpendicular to the $x^3$-axis.
There exists a continuous family of the domain wall solutions, given by \cite{Nitta:2012wi}
\beq
&& \theta^{(0)}(x^3) = \arctan \exp (\pm \sqrt 2 m_4 (x^3 -X^3))
,  \quad 0 \leq \theta^{(0)} \leq \pi , \label{eq:wall}\\
&&  n^{(0)}_i = n^{(1)}_i \sin \theta^{(0)}(x^3) , (i=1,2,3), \quad 
({\bf n}^{(1)})^2 = \sum_{ i =1}^3 (n^{(1)}_i)^2=1, \\
&& n^{(0)}_4 = \cos \theta^{(0)}(x^3).
\eeq
The width of the domain wall is $\Delta x^3 = m_4^{-1}$.
This domain wall solution has moduli ${\bf R}\times S^2$ parametrized 
by the position $X^3$ in the $x^3$-coordinate and the internal orientation ${\bf n}^{(1)}$, respectively. 
This wall is called non-Abelian since it carries non-Abelian moduli $S^2$.
The $O(3)$ symmetry of the vacua is spontaneously broken down to 
its subgroup $O(2)$ in the presence of the domain wall.
Consequently, there appear
Nambu-Goldstone modes $S^2\simeq O(3)/O(2)$ 
which are localized in the vicinity of the wall. 

Let us construct the low-energy effective theory 
on the domain wall \cite{Nitta:2012wi} 
using the moduli approximation \cite{Manton:1981mp,Eto:2006uw}:
\beq
 {\cal L}^{(1)}  =  { \sqrt 2 \over 2m_4}  (\del_{\mu} {\bf n}^{(1)})^2  
 +{T_{\rm wall} \over 2} (\del_{\mu} X^3)^2 -T_{\rm wall} , \quad 
   ({\bf n}^{(1)})^2=1 ,  \label{eq:wall-eff}
\eeq
with the domain wall tension $T_{\rm wall}= 2 \sqrt 2 m_4$. 
Here $\mu$ runs $0,1,2$. 
Hereafter, we use the same letter $\mu$ for labeling indices of 
various dimensions unless a confusion exists.
We have seen that the above $S^2$ zero modes are 
in fact normalizable modes.

By turning on the small mass $m_3(\ll m_4)$ perturbatively, 
we can calculate the effect on the domain wall theory as 
the following effective potential: 
\beq
V^{(1)}
= - {\sqrt 2 \over m_4} m_3^2 (n^{(1)}_{3})^2 + {\rm const.} 
  \label{eq:pot3}
\eeq
With ignoring the position modulus $X^3$,
the effective Lagrangian is summarized as
\beq
 &&{\cal L}^{(1)}  =  { \sqrt 2 \over m_4}  
 \left[\1{2} (\del_{\mu} {\bf n}^{(1)})^2 
      - m_3^2\left(1-  (n^{(1)}_3)^2\right)  \right] + {\rm const.} \label{eq:wall-eff2}
\eeq
We thus have obtained an $O(3)$ sigma model with 
the potential term admitting
two discrete vacua, given by $n^{(1)}_3=\pm 1$.
This model is equivalent to the massive ${\bf C}P^1$ model \cite{Abraham:1992vb}.

There exists a domain wall solution interpolating between 
the two discrete vacua $n^{(1)}_3=\pm 1$ \cite{Abraham:1992vb,Arai:2002xa}
in the wall effective theory obtained in Eq.~(\ref{eq:wall-eff2}). 
The domain wall solution perpendicular to the $x^2$-coordinate is obtained as
\beq
&& \theta^{(1)}(x^2) = \arctan \exp (\pm \sqrt 2 m_3 (x^2 -X^2)),
  \quad 0 \leq \theta^{(1)} \leq \pi,  \label{eq:wall2}\\
&&  n^{(1)}_i = n^{(2)}_i \sin \theta^{(1)}(x^2) , (i=1,2), \quad 
({\bf n}^{(2)})^2 = \sum_{i =1}^2 (n^{(2)}_i)^2=1, \\
&& n^{(1)}_3 =  \cos \theta^{(1)}(x^2).
\eeq
The width of the domain wall is $\Delta x^2 = m_3^{-1}$. 
Here, the index $i$ runs $1,2$; we also use the same letter $i$ for labeling different dimensional vectors unless a confusion exists.
This domain wall solution has moduli ${\bf R}\times S^1$ parametrized 
by the position $X^2$ in the $x^2$-coordinate and the internal orientation ${\bf n}^{(2)}$, respectively. 
Then, the total configuration is a domain wall inside a domain wall 
as schematically shown in Fig.~\ref{fig:wall-wall}(b). 
This configuration was studied before in Ref.~\cite{Ritz:2004mp}. 
The second domain wall does not have topological meaning in the bulk 
outside the first domain wall.
A wall within a wall has been also studied in different models \cite{Eto:2005sw,Auzzi:2006ju,Nitta:2012xq}.

Before going to the next subsection, let us make a comment. 
If one makes a loop of the second domain wall (inside the first domain wall), 
it is of course unstable to decay.
One can twist the $U(1)$ modulus of the domain wall 
when making a domain wall loop. 
This twisted domain wall loop carries a lump charge 
in $d=2+1$, which implies 3D Skyrme charge as discussed in \cite{Nitta:2012wi}. 
Such the twisted domain wall loop may be still unstable against shrinking. 
In order to make it stable, one can give a linear time dependence 
on the $U(1)$ modulus of the domain wall. 
Then, such a time-dependent, twisted domain wall loop 
is nothing but a Q-lump in $d=2+1$ dimensions \cite{Leese:1991hr} inside a domain wall, 
which was studied in \cite{Nitta:2012wi}.
This corresponds to a spinning 3D Skyrmion in the bulk.

\subsection{3D Skyrmion from domain walls}
Here, we construct a 3D Skyrmion in $d=3+1$ dimensions 
as a composite state of domain walls.  
To this end, let us consider the following potential term  
instead of the potential given in Eq.~(\ref{eq:pot34}): 
\beq 
V({\bf n}) = m_4^2 \left(1-({n^{(0)}_4})^2\right)  
+ m_3^2 \left(1-({n^{(0)}_3})^2\right)  
+ m_2^2 \left(1-n^{(0)}_2 \right), \quad 
  m_2 \ll m_3 \ll m_4, \label{eq:pot234}
\eeq
with a hierarchical breaking
\beq
 O(4) \stackrel{m_4}{\to} O(3) \stackrel{m_3}{\to} O(2)  \stackrel{m_2}{\to} \{0\}.
\eeq 
The difference with the previous potential in Eq.~(\ref{eq:pot34}) is the last term with $m_2$. 
Note that we take this term linear with respect to the field $n^{(0)}_2$ 
unlike the other two terms, which are quadratic with respect to the fields. 
The reason to consider this term is clarified below. 

First, we place a domain wall perpendicular to the $x^3$-axis, 
which interpolates the vacua $n^{(0)}_4 =\pm 1$ as before.
Except for the position modulus, 
a potential term due to $m_2$ is induced on 
the domain wall world-volume effective theory given in Eq.~(\ref{eq:wall-eff2}):
\beq
 &&{\cal L}^{(1)}  =  { \sqrt 2 \over m_4}  
 \left[\1{2} (\del_{\mu} {\bf n}^{(1)})^2 
      + \left(1-m_3^2  (n^{(1)}_3)^2\right)  \right] 
   - {\sqrt 2 \pi \over 2m_4} \left(1- m_2^2 n^{(1)}_2 \right)
   + {\rm const.} ,\label{eq:wall-eff3}
\eeq
with $\mu=0,1,2$.

Second, we consider the second domain wall perpendicular to the $x^2$-axis 
inside the first domain wall, 
interpolating between the vacua $n^{(1)}_3 = \pm 1$ 
in the regime of $m_2 \ll m_3$.
We can repeat the same procedure to obtain 
the effective theory on the wall inside the wall:  
\beq
 {\cal L}^{(2)}  &=&   { \sqrt 2 \over m_3}  { \sqrt 2 \over m_4}  
    \1{2} (\del_{\mu} {\bf n}^{(2)})^2 
 - {\sqrt 2 \pi \over 2m_3} {\sqrt 2 \pi \over 2m_4}
\left(1-m_2^2 n^{(2)}_2\right) 
   + {\rm const.} , 
\label{eq:wall-eff4}
\eeq
with $\mu=0,1$ and
${\bf n}^{(2)} = (n^{(2)}_1,  n^{(2)}_2 )$ 
with the constraint $(n^{(2)}_1)^2+ (n^{(2)}_2)^2=1$.
This is precisely the sine-Gordon model:
\beq
 {\cal L}^{(2)}  
 = {2\over m_3 m_4} \left[ \1{2} (\del_{\mu} \theta^{(2)})^2 
 + \hat m_2^2 \sin^2 \theta^{(2)} \right] + {\rm const.} ,
\eeq
with $(n^{(2)}_1,  n^{(2)}_2 ) = (\sin\theta^{(2)},\cos \theta^{(2)})$ 
($0 \leq \theta^{(2)} < 2\pi$)
and $\hat m_2 \equiv {\pi \over 2}m_2$.

\begin{figure}[ht]
\begin{center}
\includegraphics[width=0.7\linewidth,keepaspectratio]{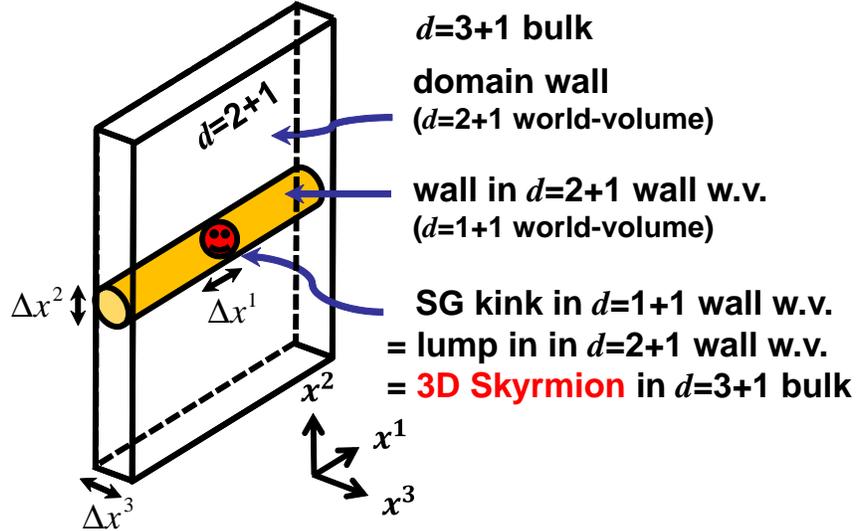}
\caption{
A 3D Skyrmion as a sine-Gordon kink inside a wall within a wall.
\label{fig:kink-wall-wall}}
\end{center}
\end{figure}

Finally, we construct a sine-Gordon kink in the $x^1$-coordinate 
as the third domain wall:
\beq
&& \theta^{(2)}(x^1) = \arctan \exp (\pm \sqrt 2 \hat m_2 (x^1 -X^1)),
 \label{eq:wall3}
\eeq
with the position modulus $X^1$ in the $x^1$-coordinate.
The total configuration is the sine-Gordon kink inside the domain wall within the domain wall,  
as drawn in Fig.~\ref{fig:kink-wall-wall}.  

In $d=2+1$ dimensional world-volume of the first domain wall, 
this sine-Gordon kink is a lump (2D Skyrmion) 
with the unit charge of the lump charge $\pi_2(S^2) \simeq {\bf Z}$ \cite{Nitta:2012xq}.  
Once it is identified with a lump in $d=2+1$, 
one can conclude from the result in \cite{Nitta:2012wi} 
that the sine-Gordon kink precisely corresponds to the 3D Skyrmion with 
the unit charge of the 3D Skyrmion charge $\pi_3(S^3) \simeq {\bf Z}$ in $d=3+1$ dimensional bulk.
We thus have constructed a 3D Skyrmion as a sine-Gordon kink inside 
a ${\bf C}P^1$ domain wall with a modulus $U(1)$ inside a non-Abelian domain wall 
with moduli $S^2$.

If we consider the quadratic mass term $m_2^2 \left(1-(n^{(0)}_2)^2\right)$ 
instead of the linear term in Eq.~(\ref{eq:pot234}), 
we have a Skyrmion with a half charge in $\pi_3(S^3)$ as the minimum soliton 
instead of a Skyrmion with the unit charge. 
The unit charge Skyrmion is split into two half charge Skyrmions.
Therefore, we have considered the linear mass 
term for $n^{(0)}_2$ although the quadratic mass term for the rest fields. 

Note that we did not need a Skyrme term. 
Therefore this type of a 3D Skyrmion can exist in principle in condensed matter physics.

If we introduce the Skyrme term on the other hand, 
a 3D Skyrmion can stably exist in the bulk. 
However it will be absorbed into the first domain wall, 
subsequently it will be absorbed into the second domain wall, and becomes a sine-Gordon kink 
as discussed in this section.

\section{Higher Dimensional Skyrmions} \label{sec:Ndim}
We consider an $O(N+1)$ model in $N$ dimensional space 
($d = N+1$ dimensional space-time) 
with the target space $S^{N}\simeq O(N+1)/O(N)$. 
Here, we show that an $N$-dimensional Skyrmion can be 
constructed as a composite state of one sine-Gordon kink and $N-1$ 
non-Abelian domain walls. 

\begin{figure}[ht]
\begin{center}
\includegraphics[width=0.6\linewidth,keepaspectratio]{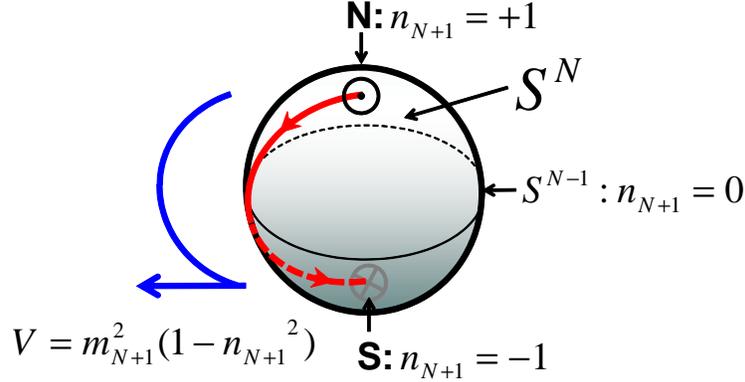}
\caption{The $S^N$ target space of an $O(N+1)$ model with a potential 
admitting two discrete vacua. 
\label{fig:Nsphere}
}
\end{center}
\end{figure}

In terms of an $O(N+1)$ vector of scalar fields 
${\bf n}^{(0)}=(n^{(0)}_1(x),\cdots,n^{(0)}_{N+1}(x))$ with the constraint 
$({\bf n}^{(0)})^2=1$, 
we consider an $O(N+1)$ model with Lagrangian 
($\mu=0,1,\cdots,N$)
\beq 
&& {\cal L}^{(0)} = \1{2} (\partial_{\mu} {\bf n}^{(0)})^2 -V({\bf n}^{(0)}),  \non
&&V^{(0)} ({\bf n}^{(0)}) = \sum_{i=3}^{N+1}m_{i}^2 \left(1-(n^{(0)}_i)^2\right)
         + m_2^2 (1-n_2^{(0)}),   \quad  
\label{eq:potential-higher}
\eeq
in $d = N+1$ space-time dimensions, 
with the hierarchical masses $m_2 \ll m_3 \ll \cdots \ll m_{N+1} $ 
and the hierarchical breaking
\beq
 O(N+1) \stackrel{m_{N+1}}{\longrightarrow} O(N) \stackrel{m_N}{\longrightarrow} \cdots 
\stackrel{m_{N-k+1}}{\longrightarrow} O(N-k) 
\stackrel{m_{N-k}}{\longrightarrow} \cdots 
\stackrel{m_3}{\longrightarrow}  O(2)  \stackrel{m_2}{\longrightarrow} \{0\}.
\eeq 
The target space is ${O(N+1) / O(N)} \simeq S^N$, see Fig.~\ref{fig:Nsphere}.

First, we consider only the largest mass $m_{N+1}$ ignoring the rests.
A domain wall solution perpendicular to the $x^{N}$-coordinate  
can be obtained as before: 
\beq
&& \theta^{(0)}(x^N) = \arctan \exp (\pm \sqrt 2 m_{N+1} (x^N -X^N))
,  \quad 0 \leq \theta^{(0)} \leq \pi , \label{eq:wall}\\
&&  n^{(0)}_i = n^{(1)}_i \sin \theta^{(0)}(x^N) , (i=1,\cdots,N), \quad 
({\bf n}^{(1)})^2 = \sum_{ i =1}^N (n^{(1)}_i)^2=1, \\
&& n^{(0)}_4 = \cos \theta^{(0)}(x^N).
\eeq
This domain wall has the moduli space ${\bf R} \times S^{N-1}$ 
parametrized by the position modulus $X^N$ and
the internal orientational moduli ${\bf n}^{(1)}$, respectively. 
We obtain an $O(N)$ model with the target space $S^{N-1}$ 
as the effective theory on the domain wall.

We then turn on the second largest mass $m_N$ perturbatively, 
which induces a potential term on the wall effective action. 
We obtain the second domain wall as before.
We repeat the same procedures. 
The effective Lagrangian on the $k$-th domain wall 
is an $O(N-k+1)$ model in ($N-k$) space dimensions [$d=N-k+1$ space-time dimensions] 
with the target space $S^{N-k}\simeq O(N-k+1)/O(N-k)$, 
described by fields ${\bf n}^{(k)} = (n^{(k)}_1(x), \cdots,n^{(k)}_{N-k+1}(x))$ 
with the constraint $({\bf n}^{(k)})^2 =1$:
\beq
 {\cal L}^{(k)}  &=&   {(\sqrt 2)^k \over \prod_{a=N-k+1}^{N+1} m_a} 
   \left[ \1{2} (\del_{\mu} {\bf n}^{(k)})^2 
 -  \sum_{i=3}^{N-k}m_{i}^2 \left(1-(n^{(k)}_i)^2\right)
\right] \non
&& -{\left( {\sqrt 2 \pi / 2}\right)^k \over \prod_{a=N-k+1}^{N+1} m_a} 
  (1-m_2^2 n^{(k)}_2 )
   + {\rm const.} \label{eq:wall-eff4}
\eeq
The ($k+1$)-th domain wall solution perpendicular to the $x^{N-k+1}$-coordinate can be obtained as
\beq
&& \theta^{(k)}(x^{N-k+1}) = \arctan \exp (\pm \sqrt 2 m_{N-k+2} (x^{N-k+1} -X^{N-k+1}))
,  \quad 0 \leq \theta^{(k)} \leq \pi,  \label{eq:wall}\\
&&  n^{(k)}_i = n^{(k+1)}_i \sin \theta^{(k)}(x^{N-k+1}) , (i=1,\cdots,N-k+1), \quad \non
&& ({\bf n}^{(k+1)})^2 = \sum_{ i =1}^{N-k} (n^{(k+1)})^2=1, \\
&& n^{(k)}_4 = \cos \theta^{(k)}(x^{N-k+1}).
\eeq
The transverse size of the ($k+1$)-th domain wall is $\Delta x^{N-k+1} = m_{N-k+2}^{-1}$. 
The ($k+1$)-th domain wall has moduli 
${\bf R} \times S^{N-k-1} \simeq {\bf R} \times O(N-k)/O(N-k-1)$. 

After repeating the same procedures $N-1$ times in total, 
we construct a sine-Gordon kink given in Eq.~(\ref{eq:wall3}) with 
$\hat m_2 \equiv (\pi/4)^{N-1} m_2$ by using the lightest mass $m_2$.
We thus have realized an $N$-dimensional Skyrmion  
characterized by the topological charge $\pi_N(S^N) \simeq {\bf Z}$
as a sine-Gordon kink (inside a non-Abelian domain wall)$^{N-1}$. 
The moduli space of this Skyrmion is ${\bf R}^N \times S^{N-1}$.
We did not need the Skyrme term for stability 
while it is unstable against shrinkage outside the walls.

\section{Summary and Discussion \label{sec:summary} }

We have constructed an $N$ dimensional Skyrmion 
characterized by the topological charge $\pi_N(S^N) \simeq {\bf Z}$ 
as a composite state of one sine-Gordon kink 
and $N-1$ non-Abelian domain walls 
in the $O(N+1)$ nonlinear sigma model 
with the $N$ hierarchical masses  
in $d=N+1$ space-time dimensions, 
where the $k$-th domain wall is placed orthogonal 
to the $x^{N-k+1}$-coordinate and has the width $\Delta x^{N-k+1} = m_{N-k+2}^{-1}$ 
and the moduli ${\bf R} \times S^{N-k-1} $. 
We have found a matryoshka structure of Skyrmions in diverse dimension, 
as summarized in Table~\ref{fig:skyrmion-hierarchy}.

\begin{table}[ht]
\begin{center}
\includegraphics[width=0.9\linewidth,keepaspectratio]{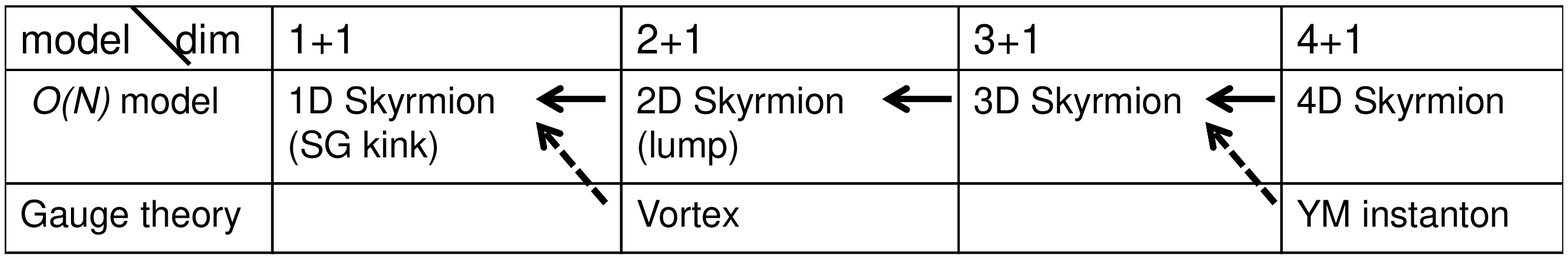}
\caption{Hierarchical or matryoshka structure of Skyrmions in diverse dimensions.
\\
An $N+1$ dimensional Skyrmion becomes an $N$ dimensional Skyrmion 
inside a non-Abelian domain wall, as indicated by arrows. 
In addition to the sequence of Skyrmions in the $O(N)$ models in different dimensions, 
two exceptional sequences from solitons in gauge theories are known, 
indicated by dotted arrows:  
a Yang-Mills instanton to a 3D Skyrmion \cite{Eto:2005cc}
and a vortex to a sine-Gordon kink \cite{Nitta:2012xq}.
\label{fig:skyrmion-hierarchy}
}
\end{center}
\end{table}

In addition to the sequence of $N$ dimensional Skyrmions in $O(N+1)$ models 
found in this paper, there exist two exceptional sequences from solitons in gauge theories known thus far, that is,  a Yang-Mills instanton to a 3D Skyrmion \cite{Eto:2005cc}
and a vortex to a sine-Gordon kink \cite{Nitta:2012xq}.
In the former, Yang-Mills instanton becomes a 3D Skyrmion 
\cite{Eto:2005cc} once it is placed inside 
a non-Abelian domain wall with $U(2)$ moduli \cite{Shifman:2003uh}, 
which explains a construction of 3D Skyrmion from instanton holonomy 
\cite{Atiyah:1989dq}. 
In the latter, a vortex becomes a sine-Gordon kink \cite{Nitta:2012xq} 
once it is absorbed into a ${\bf C}P^1$ domain wall with a $U(1)$ modulus 
\cite{Abraham:1992vb,Arai:2002xa},  
which explains a construction of a sine-Gordon kink from 
a vortex holonomy \cite{Sutcliffe:1992ep}. 
In $d=2+1$, lumps can be obtained in the strong gauge coupling limit 
of semi-local vortices. 
In $d=4+1$, we do not know whether there exists a certain relation 
between Yang-Mills instantons and 4D Skyrmions.

\section*{Acknowledgements}

This work is supported in part by 
Grant-in-Aid for Scientific Research (No. 23740198) 
and by the ``Topological Quantum Phenomena'' 
Grant-in-Aid for Scientific Research 
on Innovative Areas (No. 23103515)  
from the Ministry of Education, Culture, Sports, Science and Technology 
(MEXT) of Japan.


\newcommand{\J}[4]{{\sl #1} {\bf #2} (#3) #4}
\newcommand{\andJ}[3]{{\bf #1} (#2) #3}
\newcommand{\AP}{Ann.\ Phys.\ (N.Y.)}
\newcommand{\MPL}{Mod.\ Phys.\ Lett.}
\newcommand{\NP}{Nucl.\ Phys.}
\newcommand{\PL}{Phys.\ Lett.}
\newcommand{\PR}{ Phys.\ Rev.}
\newcommand{\PRL}{Phys.\ Rev.\ Lett.}
\newcommand{\PTP}{Prog.\ Theor.\ Phys.}
\newcommand{\hep}[1]{{\tt hep-th/{#1}}}

\end{document}